\documentclass[conference]{IEEEtran}
\IEEEoverridecommandlockouts

\usepackage[T1]{fontenc}
\usepackage{amsmath,amssymb}
\usepackage{graphicx}
\usepackage{booktabs}
\usepackage{array}
\usepackage{cite}
\usepackage{tikz}
\usetikzlibrary{arrows.meta,positioning,calc}
\usepackage{xcolor}
\usepackage{balance}
\usepackage{url}

\graphicspath{{figures/}}

\title{EMT Assessment of SCR and X/R Effects on Current-Priority-Dependent LVRT Response of a Grid-Following BESS}

\author{%
\IEEEauthorblockN{Dileep Cheerla, Amirthagunaraj Yogarathnam, Divya Jain, and Satish Natti}
\IEEEauthorblockA{GridAxon, Inc., USA\\
Email: \{dileep, araj, divya, satish\}@gridaxon.io}}

\begin{document}
\maketitle
\begin{center}
\vspace{-0.7em}
\footnotesize\itshape
Accepted for presentation and publication at the 2027 IEEE Texas Power and Energy Conference (TPEC 2027).\\
This author-prepared manuscript may differ from the final version published in IEEE Xplore.
\vspace{-0.7em}
\end{center}

\begin{abstract}
Low-voltage ride-through (LVRT) forces a grid-following inverter to allocate finite converter current between active- and reactive-current objectives. This paper uses a PSCAD electromagnetic-transient (EMT) representation of an aggregated battery energy storage system (BESS) to quantify how short-circuit ratio (SCR) and grid impedance X/R ratio alter that tradeoff. Active-current priority and reactive-current priority are compared for five paired grid conditions under the same 150-ms balanced three-phase fault, with pre-fault active power approximately 0.84 pu, POI reactive-power exchange near 0 pu, and a 1.0-pu dq-current limit. Fundamental positive-sequence voltage and power are reconstructed from 100-$\mu$s three-phase EMT outputs. Across the fixed-fault SCR sweep, the Q-priority late-fault voltage advantage increases from 0.046 pu in the strongest tested grid to 0.129 pu in the weakest; the weakest P-priority case also shows active-power reversal and a pronounced controller-frequency excursion. At fixed SCR, changing X/R produces a smaller increase from 0.092 to 0.112 pu, mainly through reduced P-priority voltage support. Because the common physical fault also deepens the passive retained-voltage condition as SCR decreases, the SCR result is interpreted as a fixed-fault grid-strength sensitivity rather than a constant-sag comparison. Grid strength is the larger sensitivity in the tested matrix, while X/R remains an additional modifier under the studied fault conditions.
\end{abstract}


\section{Introduction}
Transmission-connected inverter-based resources (IBRs) are expected to remain connected through specified abnormal-voltage conditions and provide controlled current response during ride-through events \cite{ieee2800}. Once the converter current limit is reached, the controller must divide finite current capability between active- and reactive-current objectives. IEEE Std 2800 uses reactive-current priority as the default balanced-fault ride-through mode unless active-current priority is specified by the transmission-system owner \cite{ieee2800}; the P-priority cases considered here therefore represent the permitted alternative operating mode rather than a recommended default.

Prior phasor-domain studies have shown that active- and reactive-current priority can materially change disturbance behavior of grid-following IBRs \cite{daryabak2019,fang2020}. Utility-scale BESS work has likewise demonstrated strong coupling among active power, reactive power, converter limits, and weak-grid voltage response \cite{dozein2021}. Current-limit design has received renewed attention in the context of IEEE Std 2800 ride-through requirements \cite{baker2023}. Although developed for grid-forming control, related fault-control work has implemented a Q-priority AC current-limiting structure under fault conditions \cite{cheerla2026currentlimit}; the present study instead evaluates priority selection in a grid-following BESS. In parallel, NERC guidance emphasizes EMT studies for IBR interconnections, while WECC generic BESS specifications provide a complementary planning-model framework \cite{nerc2025emt,wecc2024bess}.

EMT studies retain fast voltage, current, synchronization, and current-limiting interactions during fault inception and clearing. Existing EMT investigations have examined current-priority behavior from protection and generic fault-contribution perspectives \cite{jafarian2024,kim2024}, while phasor-versus-EMT comparisons show that model choice can affect converter fault behavior \cite{lacerda2023}. Separately, grid-following converter behavior depends on grid strength and network impedance angle \cite{ravanji2024,li2022,huang2025,wang2025}. Severe voltage depressions can also challenge synchronization of PLL-based grid-following converters when substantial current is maintained during a fault \cite{goksu2014,taul2019}.

The combined influence of SCR and X/R on the LVRT tradeoff between active- and reactive-current priority has received less direct attention. This paper addresses that intersection using paired EMT simulations of an aggregated grid-following BESS. The main contributions are: 1) matched P-priority/Q-priority comparisons at approximately 0.84 pu active power and zero pre-fault POI reactive-power exchange; 2) separate SCR and X/R sweeps under one common physical fault, with the SCR sweep interpreted explicitly as a fixed-fault rather than constant-sag sensitivity; and 3) direct dq-current evidence and raw-waveform-derived positive-sequence voltage/power metrics linking current allocation to POI response, supplemented by electrical angle and controller-frequency evidence in the weakest tested P-priority case.

\section{Aggregated EMT Model and Current-Priority Control}
\subsection{Study System}
\begin{figure*}[t]
\centering
\resizebox{0.94\textwidth}{!}{%
\begin{tikzpicture}[
  font=\small,
  eblock/.style={draw, rounded corners=2pt, align=center, minimum height=10mm, minimum width=21mm},
  cblock/.style={draw, rounded corners=2pt, align=center, minimum height=9mm, minimum width=26mm, fill=gray!5},
  note/.style={align=center, font=\footnotesize},
  line/.style={-Latex, line width=0.8pt},
  node distance=7mm and 8mm]

\node[eblock] (dc) {BESS\\DC source};
\node[eblock, right=of dc] (conv) {Aggregated GFL\\converter};
\node[eblock, right=of conv] (xf1) {LV/MV\\transformer};
\node[eblock, right=of xf1] (coll) {Equivalent\\collection system};
\node[eblock, right=of coll] (xf2) {Main\\transformer};
\node[eblock, right=of xf2, minimum width=14mm] (poi) {POI};
\node[eblock, right=of poi, minimum width=22mm] (zth) {$Z_{\mathrm{th}}$\\$R_{\mathrm{th}}+jX_{\mathrm{th}}$};
\node[eblock, right=of zth] (grid) {Thevenin\\grid};

\foreach \a/\b in {dc/conv,conv/xf1,xf1/coll,coll/xf2,xf2/poi,poi/zth,zth/grid}
  \draw[line] (\a)--(\b);

\node[note, above=10mm of poi] (fault) {Balanced 3$\phi$-G fault at POI\\$t=5.00$--$5.15$ s};
\draw[line] (fault)--(poi);
\node[note, above=10mm of zth, xshift=12mm] (gridnote) {SCR sets $|Z_{\mathrm{th}}|$; X/R sets impedance angle};

\node[cblock, below=24mm of conv, xshift=-6mm] (refs) {$P_{\mathrm{cmd}},Q_{\mathrm{cmd}}$\\reference inputs};
\node[cblock, right=of refs] (lvrt) {LVRT current request\\$\Delta i_q=K_{\mathrm{LVRT}}\Delta V$};
\node[cblock, right=of lvrt, minimum width=32mm] (pri) {P/Q priority + dq limiter\\$\sqrt{(i_d^*)^2+(i_q^*)^2}\le1$ pu};
\node[cblock, right=of pri] (inner) {inner dq-current\\control};

\foreach \a/\b in {refs/lvrt,lvrt/pri,pri/inner}
  \draw[line] (\a)--(\b);
\draw[line, -] (inner.north) -- ++(0,6mm) -| ($(conv.south)+(0,-5mm)$);
\draw[line] ($(conv.south)+(0,-5mm)$) -- (conv.south);
\end{tikzpicture}}
\caption{Aggregated grid-following BESS EMT test system and LVRT current-priority control framework.}
\label{fig:architecture}
\end{figure*}
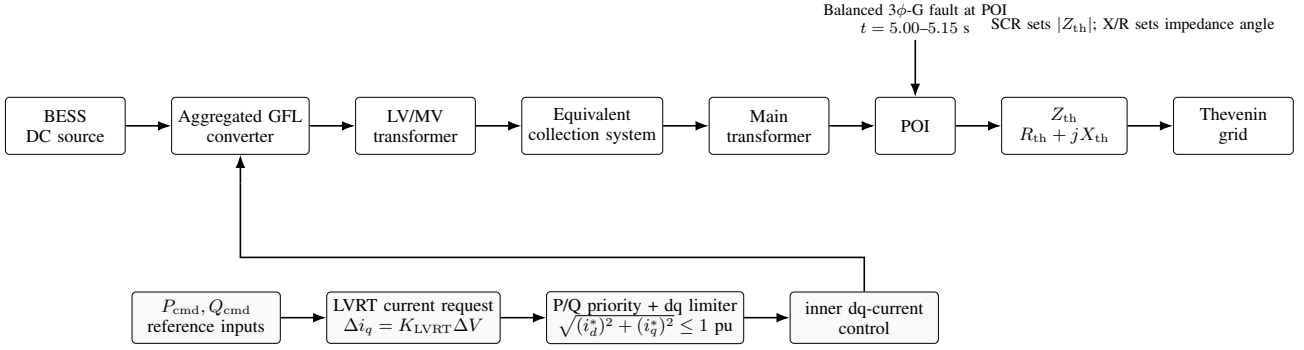

An aggregated grid-following BESS EMT model with LVRT current-priority control is implemented in PSCAD and connected to a Thevenin grid through equivalent transformation, collection-system impedance, and a main transformer. For each grid condition, pre-fault active- and reactive-power reference commands establish the operating point. The disturbance response is governed by inverter-level P/Q control, synchronization, LVRT logic, and current limiting retained in the EMT representation. Figure~\ref{fig:architecture} summarizes the electrical path and control functions relevant to the study.

Power quantities are reported in per unit on the aggregate BESS rating $S_n$; voltage quantities are reported in per unit on their corresponding nominal electrical-level bases; and dq currents use the aggregate converter rated-current base.

\subsection{LVRT Current Request and Current Limit}
Let $i_{d0}$ and $i_{q0}$ denote unconstrained active- and reactive-current requests in the synchronized dq reference frame. LVRT is asserted when inverter-terminal voltage falls below an activation threshold $V_{\mathrm{th,LVRT}}$. In the undervoltage region, the tested implementation forms a reactive-current increment with magnitude
\begin{equation}
|\Delta i_q|=K_{\mathrm{LVRT}}\left(V_{\mathrm{set,LVRT}}-V_{\mathrm{inv}}\right),
\label{eq:lvrtgain}
\end{equation}
for $V_{\mathrm{inv}}<V_{\mathrm{set,LVRT}}$, subject to the selected priority rule and converter current limit. The model sign convention uses negative $i_q$ for positive reactive-power injection, so the LVRT increment is applied in the negative-q direction. Because the activation threshold and current-law setpoint are distinct settings, a finite reactive-current increment exists at assertion. The activation threshold, hysteresis, gain, and current-law setpoint are held fixed across all simulations and are not varied in this study; they are not presented as IEEE Std 2800 prescribed values. Across all ten late-fault simulations, the resulting commanded dq-current magnitude is approximately 1.0 pu, so the current limit is active and the selected priority rule determines the d/q allocation.

The commanded-current magnitude is
\begin{equation}
I_{\mathrm{cmd}}=\sqrt{(i_d^*)^2+(i_q^*)^2}\le I_{\max},
\label{eq:currentcircle}
\end{equation}
with $I_{\max}=1.0$ pu on the aggregate converter rated-current base. A generic current-circle description of active-current priority is
\begin{align}
i_d^* &= \mathrm{sat}(i_{d0},-I_{\max},I_{\max}),\\
|i_q^*| &\le \sqrt{I_{\max}^2-(i_d^*)^2},
\label{eq:ppriority}
\end{align}
whereas reactive-current priority reverses the allocation order,
\begin{align}
i_q^* &= \mathrm{sat}(i_{q0},-I_{\max},I_{\max}),\\
|i_d^*| &\le \sqrt{I_{\max}^2-(i_q^*)^2}.
\label{eq:qpriority}
\end{align}
These relations summarize the priority ordering used to interpret the EMT response; they are not intended to reproduce implementation-specific internal logic. All non-priority inverter-control settings are held fixed within each paired comparison. Synchronization, inner-current-loop, and post-fault reference-recovery tuning are held fixed across all cases and are not varied in this study.

\subsection{Why SCR and X/R Matter}
The Thevenin grid is represented by $Z_{\mathrm{th}}=R_{\mathrm{th}}+jX_{\mathrm{th}}$. A first-order voltage-coupling relation, used only for physical interpretation, is
\begin{equation}
\Delta V \approx \frac{R_{\mathrm{th}}\Delta P+X_{\mathrm{th}}\Delta Q}{V}.
\label{eq:voltagecoupling}
\end{equation}
Thus, decreasing SCR increases network impedance magnitude and increases voltage sensitivity to converter current. At fixed impedance magnitude, increasing X/R reduces the relative voltage leverage associated with active-power transfer through $R_{\mathrm{th}}$ and increases the relative contribution associated with reactive power through $X_{\mathrm{th}}$. The nonlinear EMT simulations, not (\ref{eq:voltagecoupling}), produce the reported results \cite{ravanji2024,wang2025}.

\section{Study Methodology}
\subsection{Grid Representation and Test Matrix}
Grid strength is parameterized by
\begin{equation}
\mathrm{SCR}=\frac{S_{\mathrm{sc}}}{S_n},
\label{eq:scr}
\end{equation}
where $S_{\mathrm{sc}}$ is the three-phase short-circuit MVA and $S_n$ is the aggregate BESS rating. For $k=X/R$ and specified $|Z_{\mathrm{th}}|$,
\begin{equation}
R_{\mathrm{th}}=\frac{|Z_{\mathrm{th}}|}{\sqrt{1+k^2}},\qquad X_{\mathrm{th}}=kR_{\mathrm{th}}.
\label{eq:rx}
\end{equation}
The SCR sweep uses SCR=5, 3, and 2 at X/R=8. The X/R sweep uses X/R=3, 5, and 8 at SCR=3, so $|Z_{\mathrm{th}}|$ remains fixed in that sweep. The SCR=3, X/R=8 pair is shared.

A balanced three-phase-to-ground fault is applied at the POI through the same physical impedance in every simulation. On the selected per-unit bases, the fault magnitude is $|Z_f|=0.125$ pu with fault X/R$\approx8$ at 60 Hz, and $|Z_{\mathrm{th}}|=1/\mathrm{SCR}$ pu.

For interpretation only, the known external faulted network is $Z_{\mathrm{th}}\parallel Z_f$. At X/R=8, its magnitude changes 0.0769$\rightarrow$0.1000 pu from SCR 5$\rightarrow$2, while the passive divider $|Z_f/(Z_{\mathrm{th}}+Z_f)|$ falls 0.385$\rightarrow$0.200. Thus the common physical fault becomes electrically deeper as SCR decreases; the SCR sweep is a fixed-fault, not constant-sag, sensitivity. At SCR=3, grid X/R=3$\rightarrow$8 changes the external parallel angle only about $79.8^\circ\rightarrow82.9^\circ$ because the fixed fault itself has X/R$\approx8$. These are sanity checks from existing study quantities, not PSCAD input parameters or substitutes for the EMT results.

\begin{table}[t]
\caption{Key EMT Study Parameters and Per-Unit Bases}
\label{tab:params}
\centering
\scriptsize
\setlength{\tabcolsep}{3.2pt}
\begin{tabular}{lc}
\toprule
Parameter & Value / base \\
\midrule
EMT platform & PSCAD \\
Power base & aggregate BESS rating $S_n$ \\
Voltage bases & corresponding nominal electrical-level bases \\
Active-power command & 0.84 pu on $S_n$ ($\approx84\%$ rated) \\
Reactive-power command & case-dependent pu on $S_n$ \\
dq-current base / limit & aggregate converter rated-current base / 1.0 pu \\
LVRT threshold / hysteresis & fixed across cases; not varied \\
LVRT gain / setpoint & fixed across cases; not varied \\
Network-impedance pu base & nominal POI voltage / $S_n$ \\
Fault location / type & POI / balanced 3$\phi$-G \\
Fault $|Z_f|$ / X/R & 0.125 pu / $\approx8$ \\
Fault duration & 150 ms \\
EMT solution time step & 20 $\mu$s \\
Recorded output interval & 100 $\mu$s \\
SCR values & 2, 3, 5 \\
X/R values & 3, 5, 8 \\
\bottomrule
\end{tabular}
\end{table}

\subsection{Matched Pre-Fault POI Operating Points}
The active-power command is fixed at 0.84 pu on $S_n$. For each SCR/X/R condition, the reactive-power command is adjusted on the same base so that $Q_{\mathrm{POI,pre}}\approx0$ pu, then held identical for both priority modes. Table~\ref{tab:operatingpoints} reports 4.80--4.95 s positive-sequence averages.

The Thevenin source voltage is not retuned to force $V_{\mathrm{POI,pre}}=1.0$ pu, so pre-fault POI voltage varies naturally across the tested grid conditions. Pair matching is the primary control. Normalized voltage is used as a robustness check, not as an equal-retained-voltage or constant-sag correction.

\begin{table}[t]
\caption{Matched Pre-Fault POI Operating Points}
\label{tab:operatingpoints}
\centering
\footnotesize
\setlength{\tabcolsep}{3.4pt}
\begin{tabular}{cccccc}
\toprule
SCR & X/R & $Q_{cmd}$ & $P_{1,\mathrm{pre}}$ & $Q_{1,\mathrm{pre}}$ & $V_{1,\mathrm{pre}}$ \\
 & & (pu) & (pu) & (pu) & (pu) \\
\midrule
5 & 8 & 0.097 & 0.8394 & -0.0005 & 1.0085 \\
3 & 8 & 0.100 & 0.8392 &  0.0006 & 0.9974 \\
2 & 8 & 0.108 & 0.8387 &  0.0003 & 0.9563 \\
3 & 5 & 0.095 & 0.8392 &  0.0001 & 1.0192 \\
3 & 3 & 0.090 & 0.8403 &  0.0008 & 1.0545 \\
\bottomrule
\end{tabular}
\end{table}

\subsection{Disturbance, Waveform Processing, and Metrics}
The fault is applied at $t=5.00$ s and cleared at $t=5.15$ s. All cases use the same timing; point-on-wave sensitivity is not separately varied. The EMT solution step is 20 $\mu$s and three-phase waveforms are recorded every 100 $\mu$s. A centered 167-sample ($\approx16.7$ ms) 60-Hz rectangular DFT produces RMS phase phasors and positive-sequence $V_1$ and $I_1$. Complex power is reconstructed and normalized on $S_n$ as
\begin{equation}
S_{1,\mathrm{pu}}=P_{1,\mathrm{pu}}+jQ_{1,\mathrm{pu}}=\frac{3V_1 I_1^*}{S_n}.
\label{eq:s1}
\end{equation}
Positive P/Q denotes BESS injection toward the grid. Hereafter $P_1$ and $Q_1$ denote the corresponding per-unit quantities. The fixed-frequency DFT provides a common measurement basis, although the SCR=2 P-priority off-nominal-frequency excursion can introduce additional leakage, so exact phasor values in that case are interpreted cautiously.

The principal fault-period metric is the 5.10--5.14 s average $\overline{V}_{1,f}$, with $\overline{P}_{1,f}$ and $\overline{Q}_{1,f}$ over the same window. It is not called steady state because SCR=2 P-priority remains nonstationary. The normalized voltage is
\begin{equation}
\overline{V}_{1,f,n}=\frac{\overline{V}_{1,f}}{V_{1,\mathrm{pre}}}.
\label{eq:vnorm}
\end{equation}
Recovery uses a trailing 167-sample DFT. $t_{P95}$ and $t_{V2}$ are the elapsed times from fault clearing to the start of the first 20-ms interval satisfying, respectively, $P_1\ge95\%$ of pre-fault power and $V_1$ within $\pm2\%$ of its pre-fault value. These are comparative implementation outcomes, not compliance metrics.

\section{Results}
The baseline case first establishes the current-allocation mechanism; the SCR and X/R sweeps are then compared using positive-sequence quantities reconstructed from raw EMT waveforms.

\subsection{Baseline EMT Response and Current Allocation}
\begin{figure*}[t]
\centering
\includegraphics[width=0.84\textwidth]{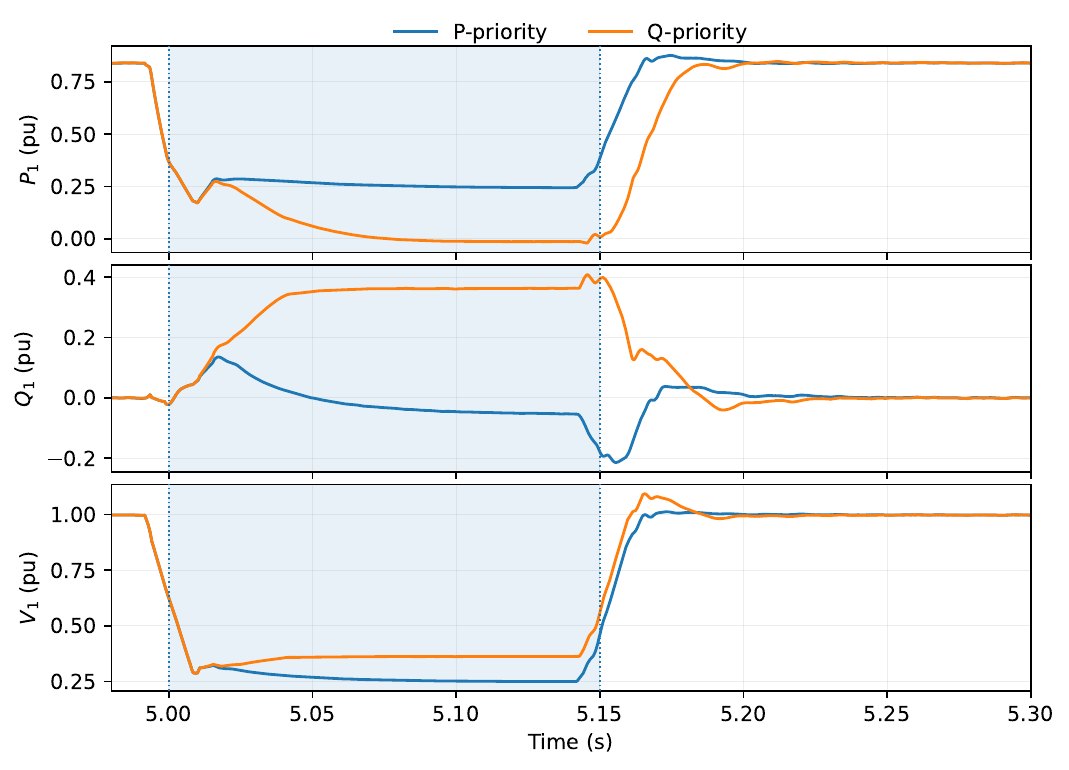}
\caption{Baseline SCR=3, X/R=8 response using positive-sequence quantities reconstructed from raw POI three-phase EMT waveforms. The shaded interval and vertical boundaries mark the physical 150-ms fault exactly from 5.00 to 5.15 s. Because $P_1$, $Q_1$, and $V_1$ are reconstructed with a centered one-cycle DFT, the displayed trajectories are smoothed across each physical fault boundary by approximately half a cycle.}
\label{fig:baseline}
\end{figure*}

The SCR=3, X/R=8 pair is the shared anchor. Figure~\ref{fig:baseline} shows the POI response. Both modes start near 0.84 pu active power, 0 pu reactive power, and 0.997 pu voltage. Over 5.10--5.14 s, P-priority gives $\overline{P}_{1,f}=0.246$ pu, $\overline{Q}_{1,f}=-0.050$ pu, and $\overline{V}_{1,f}=0.251$ pu, while Q-priority gives $-0.013$ pu, 0.363 pu, and 0.363 pu; the voltage difference is 0.112 pu. The inverter-level negative-$i_q$ command under P-priority does not imply positive POI Q because transformer, collector, and grid reactive exchange remain between the converter and POI.

\begin{figure*}[t]
\centering
\includegraphics[width=0.82\textwidth]{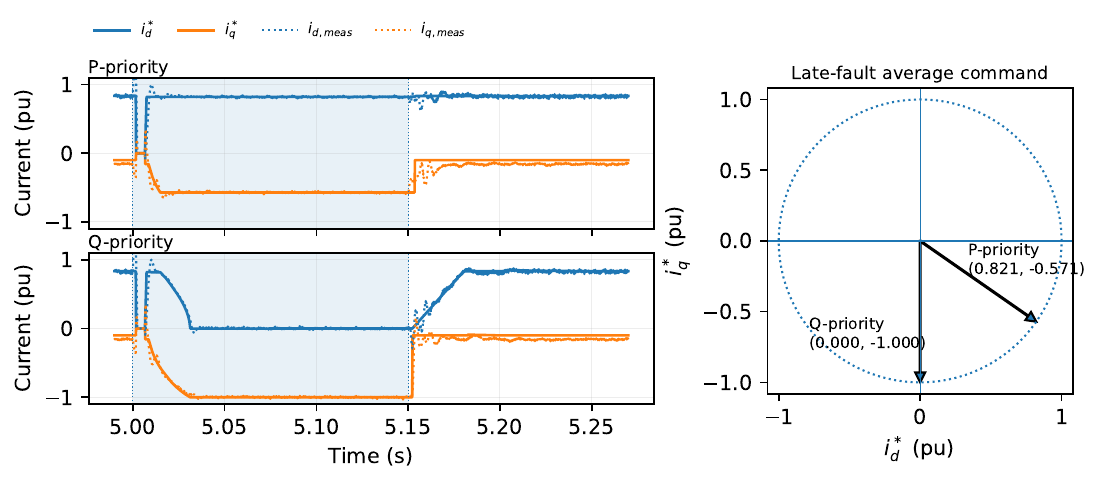}
\caption{Baseline commanded and measured dq currents and late-fault average command vectors relative to the 1.0-pu current limit.}
\label{fig:currentallocation}
\end{figure*}

Figure~\ref{fig:currentallocation} overlays commanded and measured inverter-frame dq currents. Over the late-fault window, P-priority averages $(i_d^*,i_q^*)=(0.821,-0.571)$ pu and Q-priority $(0,-1.000)$ pu; both remain on the 1.0-pu current circle and measured current magnitude is also approximately 1.0 pu. As a POI-side consistency check, $|\overline{S}_{1,f}|/\overline{V}_{1,f}\approx1.0$ pu in both baseline modes, consistent with current-limited operation. In baseline Q-priority, measured $i_q$ reaches $-0.99$ pu about 4.2 ms after the command, so the slower displayed P/Q evolution reflects evolving voltage plus the centered DFT rather than a tens-of-milliseconds current-loop lag.

A brief approximately 5-ms current-suppression transient at fault inception occurs in four of the five grid conditions in both priority modes. It ends well before the late-fault evaluation window and does not affect the reported late-fault metrics. The available recorded signals are insufficient to confirm or exclude classification of this brief event as momentary cessation.

\subsection{Effect of Grid Strength}
At X/R=8, Fig.~\ref{fig:sensitivity}(a) gives Q-priority voltage benefits of 0.046, 0.112, and 0.129 pu for SCR 5, 3, and 2; normalized differences are 0.046, 0.112, and 0.135. Because the fixed fault also deepens the passive retained-voltage condition as SCR falls, this is interpreted as a fixed-fault grid-strength trend, not a constant-sag SCR derivative. SCR=2 also contains the synchronization-related behavior discussed below.

At SCR=2 P-priority, $\overline{V}_{1,f}=0.170$ pu and $\overline{P}_{1,f}=-0.119$ pu. The MV-side reconstruction is $-0.118$ pu versus $-0.119$ pu at the POI, so collector/main-transformer losses cannot explain the sign reversal. Corresponding MV-side-to-POI differences are about 0.004 pu for baseline Q-priority and 0.004 pu for SCR2 Q-priority; because these are one-cycle positive-sequence powers during a dynamic event, they are not interpreted as pure instantaneous $I^2R$ losses.

The MV-side $\angle I_1-\angle V_1$ averages about $-124^\circ$ during the SCR2 P-priority late-fault window (roughly $-134^\circ$ to $-101^\circ$), versus about $-91^\circ$ under Q-priority. Figure~\ref{fig:sensitivity}(c) also gives $f_{\mathrm{inv}}=56.6$--61.2 Hz during the fault versus 58.6--60.5 Hz under Q-priority. Together with measured dq-current tracking and active-power reversal, these measurements show that synchronization-related dynamics materially influence P-priority. These are terminal-derived quantities rather than internal controller-frame angles, so no loss-of-synchronism threshold is claimed \cite{goksu2014,taul2019}. Q-priority maintains 0.298 pu voltage and $-0.012$ pu active power at the POI.

\subsection{Effect of X/R at Fixed SCR}
\begin{figure*}[t]
\centering
\includegraphics[width=0.74\textwidth]{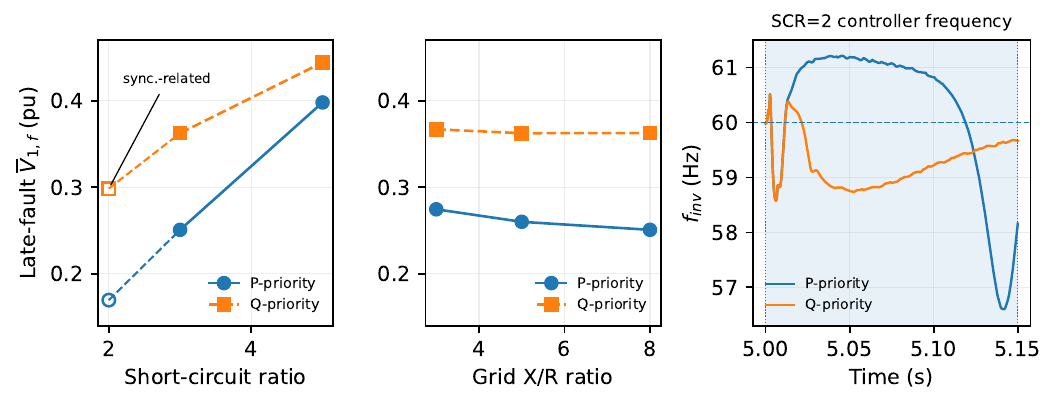}
\caption{Late-fault positive-sequence POI voltage: (a) SCR sweep at X/R=8; (b) X/R sweep at SCR=3 with fixed physical fault; and (c) SCR=2 inverter-controller frequency over the 5.00--5.15 s fault. SCR=2 is distinguished in (a) because synchronization-related dynamics also contribute.}
\label{fig:sensitivity}
\end{figure*}

At SCR=3, Fig.~\ref{fig:sensitivity}(b) gives Q-priority voltage benefits of 0.092, 0.102, and 0.112 pu for X/R=3, 5, and 8; normalized differences are 0.0876, 0.1005, and 0.1120. Q-priority voltage remains nearly constant (0.367, 0.362, 0.363 pu), while P-priority decreases (0.275, 0.260, 0.251 pu). The effect is therefore driven mainly by reduced P-priority voltage support as the network becomes more inductive. Its magnitude is smaller than the SCR-sweep change and is conditional on the tested fault, whose own X/R$\approx8$ compresses the effective external-network angle variation.

\subsection{Active-Power and Recovery Tradeoff}
Table~\ref{tab:results} gives Q-priority-minus-P-priority fault metrics and absolute recovery times. At SCR 5 and in all SCR 3 cases, Q-priority reduces late-fault active power by 0.090--0.271 pu while increasing reactive power by 0.267--0.413 pu. At SCR=2, the depressed P-priority response makes Q-priority 0.108 pu higher in active power while increasing voltage by 0.129 pu.

Q-priority consistently delays active-power recovery $t_{P95}$ by 1.9--16.5 ms across the five tested conditions, whereas the $t_{V2}$ differences are mixed and remain implementation-specific. SCR2 is the clear voltage-recovery outlier in magnitude: $t_{V2}$ is 486.8 ms for P-priority versus 54.8 ms for Q-priority, while $t_{P95}$ is 48.9 versus 54.2 ms. The long P-priority voltage recovery occurs in the same case as the large frequency/phasor excursion and is consistent with a prolonged synchronization-related transient, although the available signals do not isolate synchronization recovery as the sole cause.

\begin{table*}[!t]
\caption{Paired Fault and Recovery Metrics Across the Tested Grid Conditions}
\label{tab:results}
\centering
\scriptsize
\setlength{\tabcolsep}{6.1pt}
\begin{tabular}{c c c c c c}
\toprule
\textbf{Grid condition} & $\boldsymbol{\Delta\overline{V}_{1,f}}$ & $\boldsymbol{\Delta\overline{P}_{1,f}}$ & $\boldsymbol{\Delta\overline{Q}_{1,f}}$ & $\boldsymbol{t_{P95}}$ P/Q & $\boldsymbol{t_{V2}}$ P/Q \\
\textbf{(SCR, X/R)} & \textbf{(pu)} & \textbf{(pu)} & \textbf{(pu)} & \textbf{(ms)} & \textbf{(ms)} \\
\midrule
(5, 8) & +0.046 & -0.090 & +0.267 & 20.0 / 21.9 & 22.3 / 21.0 \\
(3, 8) & +0.112 & -0.259 & +0.413 & 21.7 / 38.2 & 22.6 / 40.2 \\
(2, 8) & +0.129 & +0.108 & +0.270 & 48.9 / 54.2 & 486.8 / 54.8 \\
(3, 5) & +0.102 & -0.271 & +0.394 & 21.6 / 37.0 & 22.4 / 39.8 \\
(3, 3) & +0.092 & -0.233 & +0.365 & 24.1 / 27.9 & 25.0 / 32.5 \\
\bottomrule
\multicolumn{6}{l}{\scriptsize $\Delta$ quantities are Q-priority minus P-priority; P/Q recovery entries denote P-priority / Q-priority.}
\end{tabular}
\end{table*}

\section{Discussion}
Two coupled mechanisms govern the response. Internally, priority selects the command vector on the fixed 1.0-pu current circle. Externally, the faulted network maps that vector into voltage and power. The SCR sweep gives the larger change in the tested matrix, but the common physical fault also produces progressively deeper passive retained voltage as SCR decreases; hence it is a fixed-fault grid-strength sensitivity, not a constant-sag comparison.

The X/R sweep holds SCR and the physical fault fixed. Q-priority voltage remains nearly constant while P-priority voltage falls as X/R increases, consistent with (\ref{eq:voltagecoupling}). The effect is smaller than the SCR-sweep change. Because the fault itself is strongly inductive, the faulted external-network angle varies over a narrower range than nominal grid X/R; the reported X/R sensitivity is therefore conditional on the tested fault characteristics.

At SCR=2, positive d-axis current does not guarantee positive active power: the MV-side current-to-voltage angle enters the absorption quadrant, $f_{\mathrm{inv}}$ has a large excursion, and active power reverses before the collector/main-transformer path. These observations are consistent with severe-fault synchronization mechanisms reported for PLL-based converters \cite{goksu2014,taul2019}. The available terminal-derived angle and frequency signals do not provide an internal controller-frame angle, so no universal instability threshold is inferred.

The study is limited to one aggregated GFL BESS implementation, one dispatch level, one balanced-fault impedance/duration and timing, non-equalized pre-fault voltage, and fixed controller settings. Detailed passive impedances within the aggregate plant representation and internal synchronization/current-loop tuning are not reported; the external-network calculation is therefore only an interpretive check using the per-unit POI-side quantities reported in the paper. The results reflect one aggregated GFL BESS implementation under fixed controller settings and should not be interpreted as representative of any specific commercial product or of grid-following BESS behavior in general. Future work should examine additional operating points, fault characteristics, controller settings, and hardware or field measurements.

\section{Conclusion}
This paper compared active- and reactive-current priority during BESS LVRT using matched pre-fault P/Q conditions, a common physical fault, the same 1.0-pu current limit, and raw-waveform-derived positive-sequence quantities. Under the fixed fault, the Q-priority late-fault voltage benefit increases from 0.046 pu in the strongest tested grid to 0.129 pu in the weakest, although the SCR=2 P-priority endpoint also includes synchronization-related active-power reversal and slow voltage recovery. At fixed SCR, increasing X/R produces a smaller but systematic change, from 0.092 to 0.112 pu, mainly through reduced P-priority voltage support; this sensitivity is conditioned by the tested inductive fault. Thus grid strength is the larger sensitivity in the studied matrix, while impedance angle remains a measurable modifier of the current-priority tradeoff. The weakest-grid synchronization and recovery behavior remains implementation-specific.

\balance
\bibliographystyle{IEEEtran}
\bibliography{references/references}

@misc{ieee2800,
  author       = {{IEEE}},
  title        = {{IEEE Std 2800-2022: IEEE Standard for Interconnection and Interoperability of Inverter-Based Resources (IBRs) Interconnecting with Associated Transmission Electric Power Systems}},
  year         = {2022},
  doi          = {10.1109/IEEESTD.2022.9762253}
}

@inproceedings{daryabak2019,
  author    = {M. Daryabak and M. Rahmatian and K. Kabiri and K. Chilukuri},
  title     = {Active Power Priority Vs. Reactive Power Priority Operating Modes in Solar Plants},
  booktitle = {2019 CIGRE Canada Conference},
  address   = {Montreal, QC, Canada},
  year      = {2019},
  note      = {Paper CIGRE-200}
}

@inproceedings{fang2020,
  author    = {F. Fang and M. Rahmatian and K. Chilukuri},
  title     = {Active and Reactive Power Priority Imposed by Current Limit Logic in Inverter-Based Generators},
  booktitle = {2020 CIGRE Canada Conference},
  address   = {Toronto, ON, Canada},
  year      = {2020},
  note      = {Paper CIGRE-295}
}

@article{dozein2021,
  author  = {M. Ghazavi Dozein and O. Gomis-Bellmunt and P. Mancarella},
  title   = {Simultaneous Provision of Dynamic Active and Reactive Power Response From Utility-Scale Battery Energy Storage Systems in Weak Grids},
  journal = {IEEE Transactions on Power Systems},
  volume  = {36}, number = {6}, pages = {5548--5557}, month = nov, year = {2021},
  doi     = {10.1109/TPWRS.2021.3076218}
}

@inproceedings{baker2023,
  author    = {W. W. Baker and M. Patel and A. Haddadi and E. Farantatos and J. C. Boemer},
  title     = {Inverter Current Limit Logic Based on the {IEEE} 2800-2022 Unbalanced Fault Response Requirements},
  booktitle = {2023 IEEE Power \& Energy Society General Meeting (PESGM)},
  address   = {Orlando, FL, USA}, pages = {1--5}, year = {2023},
  doi       = {10.1109/PESGM52003.2023.10252528}
}

@misc{cheerla2026currentlimit,
  author = {Dileep Cheerla and Amirthagunaraj Yogarathnam},
  title  = {{AC--DC} Coupled Current Limiting and Virtual Impedance-Based Control for Fault-Resilient Grid-Forming Converters},
  note   = {To be presented at \emph{IEEE Int. Conf. Commun., Control, Comput. Technol. Smart Grids (SmartGridComm)}, College Station, TX, USA, 2026}
}

@techreport{nerc2025emt,
  author      = {{North American Electric Reliability Corporation}},
  title       = {Electromagnetic Transient Studies for Interconnection of Inverter-Based Resources},
  institution = {NERC},
  type        = {Reliability Guideline},
  month       = mar,
  year        = {2025}
}

@misc{wecc2024bess,
  author = {{Western Electricity Coordinating Council}},
  title  = {Battery Energy Storage Model Spec},
  month  = aug,
  year   = {2024},
  note   = {Approved Model Specification. [Online]. Available: \url{https://www.wecc.org/wecc-document/13216}}
}

@inproceedings{jafarian2024,
  author    = {P. Jafarian and F. Aminifar},
  title     = {A Comparison of the Impact of Current Priority Modes of Inverter-Based Resources on the Performance of Grid Protective Relays},
  booktitle = {2024 International Conference on Protection and Automation in Power Systems (IPAPS)},
  address   = {Shahrood, Iran}, pages = {22--29}, year = {2024},
  doi       = {10.1109/IEEECONF62358.2024.10453157}
}

@inproceedings{kim2024,
  author    = {J. Kim and Y. Liu and B. A. Ross and Y. Chen and N. A. Samaan and E. Muljadi and S. Seo},
  title     = {Fault Contribution of Grid-Following and Grid-Forming Inverters Considering Generic Inverter Controls and Ride-Through Requirements},
  booktitle = {IECON 2024 - 50th Annual Conference of the IEEE Industrial Electronics Society},
  address   = {Chicago, IL, USA}, pages = {1--6}, year = {2024},
  doi       = {10.1109/IECON55916.2024.10905413}
}

@article{lacerda2023,
  author  = {V. A. Lacerda and E. Prieto-Araujo and Marc {Cheah-Ma\~{n}\'{e}} and O. {Gomis{-}Bellmunt}},
  title   = {Phasor and {EMT} Models of Grid-Following and Grid-Forming Converters for Short-Circuit Simulations},
  journal = {Electric Power Systems Research}, volume = {223}, pages = {109662}, year = {2023},
  doi     = {10.1016/j.epsr.2023.109662}
}

@article{ravanji2024,
  author  = {M. H. Ravanji and W. Zhou and N. Mohammed and B. Bahrani},
  title   = {Comparative Analysis of the Power Output Capabilities of Grid-Following and Grid-Forming Inverters Considering Static, Dynamic, and Thermal Limitations},
  journal = {IEEE Transactions on Power Systems}, volume = {39}, number = {2}, pages = {2693--2705}, month = mar, year = {2024},
  doi     = {10.1109/TPWRS.2023.3279373}
}

@article{li2022,
  author  = {C. Li and S. Wang and J. Liang},
  title   = {Tuning Method of a Grid-Following Converter for the Extremely-Weak-Grid Connection},
  journal = {IEEE Transactions on Power Systems}, volume = {37}, number = {4}, pages = {3169--3172}, month = jul, year = {2022},
  doi     = {10.1109/TPWRS.2022.3167899}
}

@article{huang2025,
  author  = {L. Huang and F. Blaabjerg},
  title   = {A Reference Frame Rematching Method for Enhancing Large-Signal Stability of Grid-Following Inverter Under Weak Grid},
  journal = {IEEE Transactions on Power Electronics}, volume = {40}, number = {4}, pages = {6135--6146}, year = {2025},
  doi     = {10.1109/TPEL.2024.3516214}
}

@inproceedings{wang2025,
  author    = {J. Wang and S. Ganguly and B. Kroposki},
  title     = {Experimental Evaluation of Impact of Short-Circuit Ratio ({SCR}) and {X/R} Ratio on Stability of Grid-Forming and Grid-Following Inverters},
  booktitle = {2025 IEEE Power \& Energy Society General Meeting (PESGM)},
  address   = {Austin, TX, USA}, year = {2025},
  doi       = {10.1109/PESGM52009.2025.11225025}
}

@article{goksu2014,
  author  = {O. Goksu and R. Teodorescu and C. L. Bak and F. Iov and P. C. Kjaer},
  title   = {Instability of Wind Turbine Converters During Current Injection to Low Voltage Grid Faults and {PLL} Frequency Based Stability Solution},
  journal = {IEEE Transactions on Power Systems},
  volume  = {29}, number = {4}, pages = {1683--1691}, month = jul, year = {2014},
  doi     = {10.1109/TPWRS.2013.2295261}
}

@article{taul2019,
  author  = {M. G. Taul and X. Wang and P. Davari and F. Blaabjerg},
  title   = {An Overview of Assessment Methods for Synchronization Stability of Grid-Connected Converters Under Severe Symmetrical Grid Faults},
  journal = {IEEE Transactions on Power Electronics},
  volume  = {34}, number = {10}, pages = {9655--9670}, month = oct, year = {2019},
  doi     = {10.1109/TPEL.2019.2892142}
}

\end{document}